\documentclass{article}
\usepackage{spconf}

\usepackage{subcaption} 
\usepackage{acronym}
\usepackage{bm} 
\usepackage{algorithm}
\usepackage{algpseudocode}
\usepackage{amsmath} 
\usepackage{amsfonts,graphicx,booktabs,siunitx, graphics}
\usepackage{url}
\usepackage{etoolbox}
\usepackage{comment} 
\AtBeginEnvironment{algorithm}{\tiny}

\acrodef{TF}{time-frequancy}
\acrodef{STFT}{time Fourier transform}
\acrodef{SI-SDR}{Scale-Invariant Signal-to-Distortion Ratio}
\acrodef{iSTFT}{inverse STFT}
\acrodef{RTF}{relative transfer function}
\acrodef{PC}{predictor-corrector}
\acrodef{ATFs}{Acoustic Transfer Functions}
\acrodef{RIR}{Room Impulse Response}
\acrodef{MVDR}{Minimum Variance Distortionless Response}
\acrodef{FCN}{fully convolutional network}

\title{AMDM-SE: Attention-based Multichannel Diffusion Model for Speech Enhancement }

\name{Renana Opochinsky and
      Sharon Gannot 
      }
\address{Faculty of Engineering, Bar Ilan University, Ramat-Gan, 5290002, Israel\\
\{renana.klainman,sharon.gannot\}@biu.ac.il}

\begin{document}
\ninept
\maketitle

\begin{abstract}

Diffusion models have recently achieved impressive results in reconstructing images from noisy inputs, and similar ideas have been applied to speech enhancement by treating time–frequency representations as images. With the ubiquity of multi-microphone devices, we extend state-of-the-art diffusion-based methods to exploit multichannel inputs for improved performance. Multichannel diffusion-based enhancement remains in its infancy, with prior work making limited use of advanced mechanisms such as attention for spatial modeling—a gap addressed in this paper.
We propose AMDM-SE, an Attention-based Multichannel Diffusion Model for Speech Enhancement, designed specifically for noise reduction. AMDM-SE leverages spatial inter-channel information through a novel cross-channel time–frequency attention block, enabling faithful reconstruction of fine-grained signal details within a generative diffusion framework. On the CHiME-3 benchmark, AMDM-SE outperforms both a single-channel diffusion baseline and a multichannel model without attention, as well as a strong DNN-based predictive method. Simulated-data experiments further underscore the importance of the proposed multichannel attention mechanism.
Overall, our results show that incorporating targeted multichannel attention into diffusion models substantially improves noise reduction. While multichannel diffusion-based speech enhancement is still an emerging field, our work contributes a new and complementary approach to the growing body of research in this direction.

\end{abstract}
\vspace{-0.5em}
\section{Introduction}
\label{sec:intro}



Multichannel speech enhancement has been the subject of rich and comprehensive research in model-based approaches, as reviewed in \cite{gannot2017consolidated}, where methods such as MVDR and related beamformers explicitly exploit spatial coherence. Building on these foundations, recent works have integrated deep neural networks (DNNs) with classical frameworks, yielding hybrid strategies that combine statistical modeling with data-driven learning. Representative examples include 
FaSNet \cite{luo2019fasnet}, SpatialNet \cite{quan2024spatialnet}, McNet \cite{yang2023mcnet}, and explainable multichannel architectures \cite{cohen2025explainable}, which illustrate how attention and learned beamforming mechanisms enhance separation, denoising, and dereverberation.

In parallel, diffusion models, originally developed for image synthesis \cite{ho2020denoising}, have shown remarkable success when adapted to speech. In the single-channel domain, they have been applied to enhancement~\cite{lu2022conditional}, separation~\cite{lutati2023separate}, and generation~\cite{popov2021grad}, demonstrating strong potential thanks to their ability to capture complex data distributions and generate perceptually natural audio. Unlike predictive models that directly map noisy inputs to clean targets, diffusion models iteratively refine estimates through learned denoising steps. Recent advances such as SGMSE+~\cite{richter2023speech}, StoRM~\cite{lemercier2023storm}, conditional diffusion~\cite{lu2022conditional}, and VAE-guided diffusion~\cite{yang2024pre} have further improved robustness and quality in the single-channel setting.

Nevertheless, the application of diffusion-based speech enhancement in the multichannel domain remains relatively underexplored. Recent studies have made substantial progress in this direction: the multi-stream framework \cite{nakatani2024multi} integrates model-based priors with data-driven diffusion, mSGMSE \cite{kimura2024diffusion} generalizes score-based enhancement to a MIMO architecture, and DiffCBF \cite{kimura2025diffcbf} combines convolutional beamforming with diffusion refinement. These works highlight the promise of combining physical modeling with generative methods and demonstrate the feasibility of diffusion for preserving spatial cues in multi-microphone arrays.

Building on these directions, we propose AMDM-SE, an Attention-based Multichannel Diffusion Model for Speech Enhancement. Unlike prior multichannel diffusion approaches, AMDM-SE explicitly incorporates unique cross-channel time–frequency attention, enabling the model to jointly leverage spectral, temporal, and spatial cues. Extending the SGMSE backbone \cite{richter2023speech}, our method bridges multi-microphone coherence with generative refinement. Experiments on simulated data and the  CHiME-3 dataset demonstrate that AMDM-SE outperforms both single-channel SGMSE and a multichannel variant without attention in terms of perceptual quality and intelligibility. The project page with audio samples can be found at \texttt{https://AMDMSE.github.io}
.

\begin{figure*}[t]
  \centering
  \centerline{\includegraphics[width=0.72\textwidth]{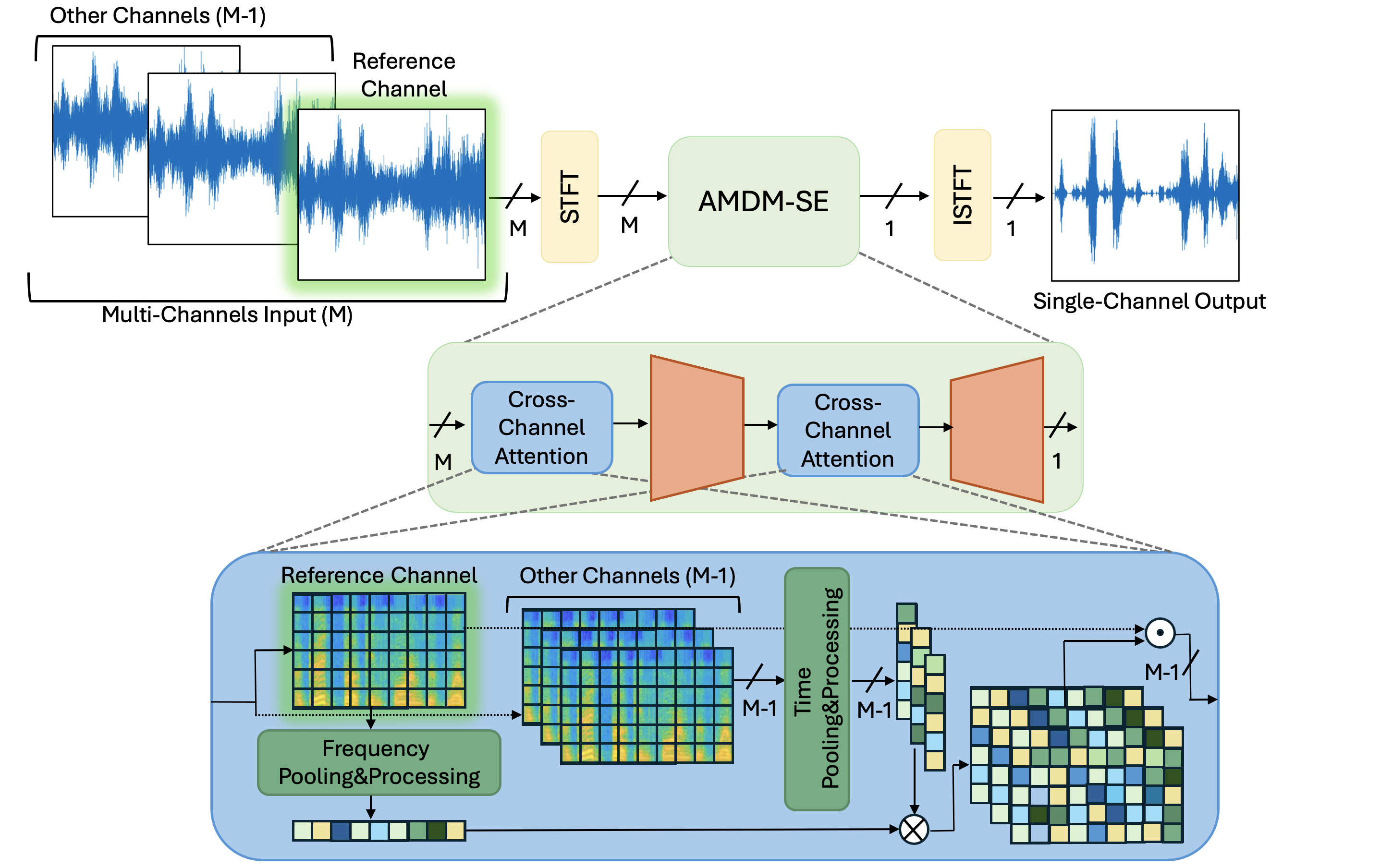}}
  \caption{Network architecture, AMDM-SE in the green frame and Cross Channel TF-Attention in the blue frame.}
  \label{fig:Network}
\end{figure*}

\vspace{-0.5em}
\section{Problem Formulation}
\label{sec:Problem}
In this paper, we address the noise reduction problem using an array of $M$ microphones. 
Let $x_m(n),\,m=1,\ldots,M$ be a noisy speech mixture recorded by the m-th microphone:
\begin{equation}
x_m(n) = h_m(n) \ast s(n) + v_m(n),
\end{equation}
where
$s(n)$ is the clean signal, $h_m(n)$ is the \ac{RIR} between the speaker and the $m$-th microphone,
and $v_m(n)$ is the additive noise component as recieved by the $m$-th microphone. 
The received microphone signal in the \ac{STFT} domain is given by $x_m(\ell,k)$, with $\ell=0,1,\dots,L-1$, $k=0,1,\ldots,K-1$, and $L,K$ are the number of time-frames and frequency bins, respectively. Define $\mathbf{x}$ as a tensor comprising all microphones, time frames, and frequency bins. Define one of the microphone channels as the reference channel $\mathbf{x}_{\text{ref}} $, and all other $M-1$ channels as $\mathbf{\bar{x}} $, such that $\mathbf{x} = \{\mathbf{x}_{\text{ref}}, \mathbf{\bar{x}}\}$. 


We opt for a complex \ac{STFT} representation as the input feature, rather than the more common magnitude-only representation. 
%
Adopting the ``engineering trick'' in \cite{richter2023speech}, we 
compress the dynamic range of the amplitudes as follows:
$x_m(\ell,k) \leftarrow \beta |x_m(\ell,k)|^{\alpha} e^{i \angle(x_m(\ell,k))}$, with $\alpha = 0.5$ and $\beta = 3$. 

The objective of the proposed method is to estimate the
clean non-reverberant speech signal $s(n)$.

\vspace{-0.5em}
\section{Method: Attention-based Multichannel Diffusion Model for Speech
Enhancement (AMDM-SE)}
\label{sec:Method}

The overall architecture of our network is illustrated in Figure~\ref{fig:Network}. The input to the network is a multichannel speech signal comprising $M$ channels. This signal is first transformed into the complex \ac{TF} domain using the \ac{STFT}. The resulting representation is then processed by the AMDM-SE module---the diffusion-based core module of our system---which produces a single-channel enhanced signal in the \ac{STFT} domain. Finally, the enhanced signal is transformed back to the time domain using the inverse STFT (iSTFT).

\subsection{AMDM-SE}

\begin{algorithm}[h]
\label{alg:AMDMSE}
\caption{AMDM-SE (following SGMSE \cite{richter2023speech})}
\begin{algorithmic}[1]
\State \textbf{Input:} Noisy spectrograms  $\mathbf{x} = \{\mathbf{x}_{\text{ref}}, \bar{\mathbf{x}} \}$, clean spectrogram $\mathbf{s}_0$
\State \textbf{Hyperparameters (Diffusion Process):} $\gamma$, $\sigma_{\min}$, $\sigma_{\max}$, $T$, $t_{\epsilon}$
\State \textbf{Output:} Enhanced spectrogram $\hat{\mathbf{s}}$

\vspace{0.5em}
\State \textbf{Forward Diffusion Process (Training)}
\For{each training sample $(\mathbf{s}_0, \mathbf{x})$}
    \State Sample $t \sim \mathcal{U}(t_{\epsilon}, T)$
    \State Compute $\boldsymbol{\mu_t} = e^{-\gamma t} \mathbf{s}_0 + (1 - e^{-\gamma t}) \mathbf{x}$
    \State Compute $\sigma_t^2 = \frac{\sigma_{\min}^2 \left( \left( \frac{\sigma_{\max}}{\sigma_{\min}} \right)^{2t} - e^{-2\gamma t} \right)}{2 \log \left( \frac{\sigma_{\max}}{\sigma_{\min}} \right)}$
    \State Sample $\mathbf{z} \sim \mathcal{N}(0, \mathbf{I})$
    \State Generate $\mathbf{s}_t = \boldsymbol{\mu_t} + \sigma_t \cdot \mathbf{z}$
    \State Compute target score: $\text{score}_\theta=\nabla_{\mathbf{s}_t} \log p(\mathbf{s}_t | \mathbf{s}_0, \mathbf{x}_{\text{ref}}, \bar{\mathbf{x}})$ 
    \State Compute loss: $\mathcal{L} = \mathbb{E} \left[ \lambda(t) \cdot \left\| \text{score}_\theta + \frac{\mathbf{s}_t - \boldsymbol{\mu_t}}{\sigma_t^2} \right\|^2 \right]$
    \State Update parameters $\theta$ using gradient descent
\EndFor

\vspace{0.5em}
\State \textbf{Reverse Sampling (Inference)}
\State Initialize $\mathbf{s}_T \sim \mathcal{N}(\boldsymbol{\mu_t}, \sigma_T^2 \mathbf{I})$
\For{$t = T$ down to $t_{\epsilon}$ with step $-\Delta t$}
    \State Compute drift: $f(\mathbf{s}_t, \mathbf{x}_{\text{ref}}) = \gamma (\mathbf{x}_{\text{ref}} - \mathbf{s}_t)$
    \State Compute diffusion: $g(t) = \sigma_{\min} \left( \frac{\sigma_{\max}}{\sigma_{\min}} \right)^t \sqrt{2 \log \left( \frac{\sigma_{\max}}{\sigma_{\min}} \right)}$
    \State Estimate score: $\text{score}_\theta(\mathbf{s}_t, t, \mathbf{x}_{\text{ref}}, \bar{\mathbf{x}})$
    \State Sample $\mathbf{z} \sim \mathcal{N}(0, \mathbf{I})$
    \State Update: 
    \Statex \quad $\mathbf{s}_{t - \Delta t} = \mathbf{s}_t + \left( f(\mathbf{s}_t, \mathbf{x}) - g^2(t) \text{score}_\theta \right) \Delta t + g(t) \sqrt{\Delta t} \cdot \mathbf{z}$
\EndFor
\State \Return $\hat{\mathbf{s}} = \mathbf{s}_0$
\end{algorithmic}
\end{algorithm}

AMDM-SE is a score-based diffusion model in the complex \ac{STFT} domain, aiming at speech enhancement using multichannel recordings. Building on SGMSE \cite{richter2023speech}, we introduce the main modifications to the original formulation and provide a concise summary of the diffusion process in Algorithm 1. 
Given the noisy and reverberant multichannel speech, $\mathbf{x}$ , 
AMDM-SE generates enhanced speech, $\hat{\mathbf{s}}$, using a conditional diffusion model. The diffusion model comprises three main stages:

\noindent\textbf{Forward Process:} Gradually adds noise to the clean spectrogram $\mathbf{s}_0$, guided by the noisy observation $\mathbf{x}$, to produce a noisy version $\mathbf{s}_t$ at step $t$, as originally introduced in \cite{song2020score} and subsequently applied to speech enhancement in \cite{richter2023speech}.

\noindent\textbf{Score Estimation:} Trains a neural network $\text{score}_\theta(\cdot)$ to approximate the gradient of the log probability density of $\mathbf{s}_t$, given $\mathbf{x}$, namely $\nabla_{\mathbf{s}_t} \log p(\mathbf{s}_t  \mathbf{s}_0, \mathbf{x}_{\text{ref}}, \bar{\mathbf{x}})$. Note that here we extend the SGMSE \cite{richter2023speech} by providing the multichannel input to the U-net rather than only a single channel in the original work and by using cross-channel \ac{TF}-attention. Explicitly, the score calculation is conditioned on all $M$ channels
. In our U-Net design, we extend the NCSN++ architecture~\cite{song2020score} to process multichannel real–imaginary STFT representations, rather than being limited to a single-channel input.

\noindent\textbf{Reverse Process:} Starting from a noisy sample $\mathbf{s}_T$, this stage iteratively denoises the sample using the estimated score function to obtain the enhanced spectrogram ${\mathbf{s}}_0$. We employ the so-called \ac{PC}
samplers proposed in \cite{song2020score}.


\subsection{Cross-Channel TF-Attention}


The Cross-Channel \ac{TF}-attention module was adopted from \cite{9966661}, where it was proposed in the context of single-channel noise reduction. The Cross-Channel \ac{TF}-attention block is depicted in Fig.~\ref{fig:Network}, in the blue frame. This block is positioned both before the diffusion network and in the bottleneck layer inside the diffusion network. The block commences with average pooling layers on the frequency dimension of the reference channel and average pooling layers on the time dimension of each of other $M-1$ channels, followed by $1 \times 1 $ Conv layers and activation functions. The Pooling and Processing blocks described in Fig.~\ref{fig:Network} consists of the following operations:
\[
\text{Average Pooling} \rightarrow 1 \times 1 \text{ Conv} \rightarrow \text{ReLU} \rightarrow 1 \times 1 \text{ Conv} \rightarrow \text{Sigmoid}
\] 
These layers produce $M-1$ attention masks, each of which is applied element-wise to the modified spectrogram of the reference channel, yielding $M-1$ masked representations of the reference channel---each conditioned on one of the remaining channels. We concatenate the $M-1$ channel output of the block with the original $M$ channels, resulting in $2M-1$ output channels. 

\section{Experimental Results}
\label{sec:Experiments}

\subsection{Datasets}

We evaluate the proposed method on both a simulated multichannel dataset and the CHiME-3 corpus, in order to assess its performance across controlled and real-world noisy acoustic conditions.

\subsubsection{Simulated Datasets}

To generate our simulated dataset, we selected clean speech utterances ranging from 4 to 9 seconds in duration from the LibriSpeech corpus \cite{panayotov2015librispeech}. Specifically, the train-clean-360 and train-clean-100 subsets were used for training, while the test-clean subset was reserved for evaluation. There is no speaker overlap between the training and test sets.

The simulated microphone array consists of four microphones arranged in a linear configuration with inter-microphone distances of 8, 6, and 8 cm. Each utterance was convolved with \ac{RIR}s generated using the image method \cite{allen1979image}, as implemented by \cite{habets2006room}. We selected a fixed and low reverberation level, setting the reverberation time to $RT_{60} = 0.2$.

The reverberated signals were further corrupted with babble noise sampled from the WHAM! dataset \cite{Wichern2019wham}, with the signal-to-noise ratio (SNR) uniformly drawn from the range [5, 15] dB. Speaker positions were randomly assigned (within physical constraints) inside shoebox-shaped rooms, whose dimensions were also randomly sampled: length and width from a uniform distribution in the range [4.5, 6.5] meters, and height from [2.5, 3] meters.

In total, the simulated dataset comprises 8000, 2000, and 2000 utterances (each 10 seconds long) for the training, validation, and test sets, respectively. The ground-truth target for each utterance is the clean, anechoic version of the signal with an alignment matching that of the reference channel (channel \#0), enabling the model to learn both denoising and dereverberation under low-reverberation conditions.

\subsubsection{CHiME-3 Dataset}
We employed the publicly available CHiME-3 dataset \cite{barker2015third}, which comprises recordings captured in four real-world noisy environments: a cafeteria, a bus, a pedestrian area, and next to a
busy street. This dataset is widely used for evaluating multichannel speech enhancement algorithms. The recordings were made using a 6-channel microphone array ($C = 6$) mounted on a tablet held by the speaker, with a sampling rate of 16 kHz.

The dataset includes 7138 training, 1640 development, and 1320 test utterances, with an average duration of approximately 3 seconds. We adopt the close-talk microphone (microphone \#0) as the ground-truth clean signal, unlike many prior works that use the reverberant 5th channel as the reference for the speech enhancement target, thus providing a more accurate target for both denoising and dereverberation. As noted in \cite{vincent2017analysis}, the CHiME-3 dataset includes certain mismatches that have led many prior studies to create modified versions of the data, making direct performance comparisons unreliable.

\subsection{Evaluation Metrics}

We assess the signal quality using PESQ \cite{rix2001perceptual}, eSTOI \cite{jensen2016algorithm} for speech intelligibility, and deep noise suppression mean opinion score (DNSMOS) \cite{reddy2022dnsmos} for overall speech quality. Additionaly, on the simulated data we also provide
 \ac{SI-SDR} measure \cite{li2020si}.

\subsection{Training Configuration}
The AdamW optimizer \cite{loshchilov2017decoupled} was used with a batch size of 24. The network was trained using early stopping based on improvements in validation PESQ. Most of the network’s training parameters follow those of the SGMSE \cite{richter2023speech}.

We chose the following diffusion parameters: $\gamma=1.5$, $\sigma_{\min}=0.05$, $\sigma_{\max}=0.5$, $T=30$, $t_{\epsilon}=0.03$ as in SGMSE.
\begin{figure*}[t]
  \centering
  \centerline{\includegraphics[width=0.7\textwidth]{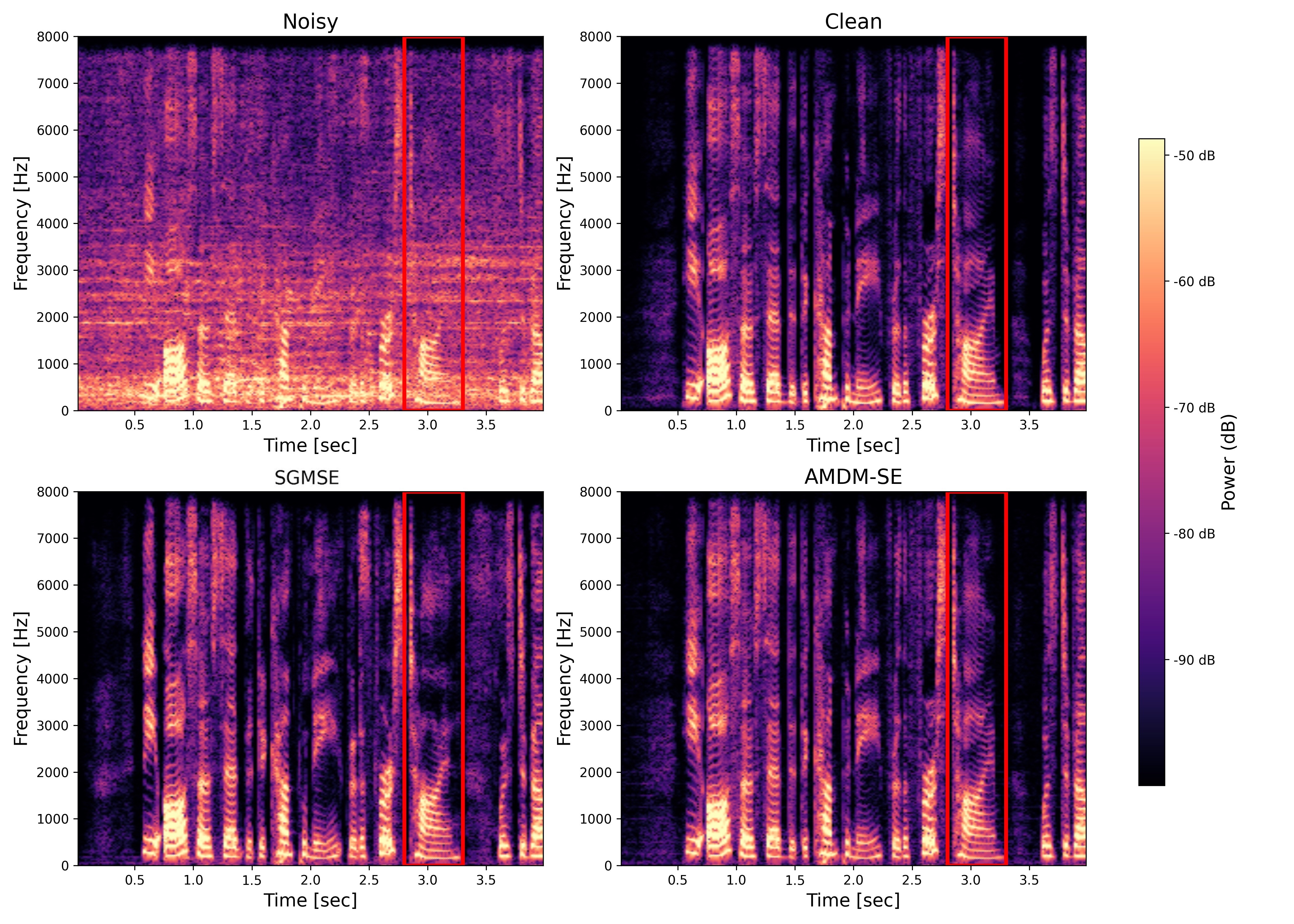}}
  \caption{Example of spectrogram of upper left: noisy, upper right: clean signal, bottom left: SGMSE output,  bottom rigth: AMDM-SE output.}
  \label{fig:spectrograms}
\end{figure*}

\subsection{Baseline Methods}

Our evaluation compares the proposed AMDM-SE with other algorithms:
1) \textbf{SGMSE} \cite{richter2023speech}, as described in Section~\ref{sec:intro}, a single-channel diffusion network; 2) \textbf{MC-SGMSE} multichannel variant of SGME (No Attention): Our multichannel extension of the SGME model, without incorporating attention mechanisms; and 3) \textbf{CA Dense U-Net} \cite{tolooshams2020channel} extends the traditional U-Net architecture with dense connections and channel attention mechanisms, enabling effective multichannel speech enhancement.
%


To the best of our knowledge, existing multichannel diffusion methods \cite{kimura2024diffusion, nakatani2024multi} have not released code or weights, and rely on different datasets, making direct comparison infeasible. 
We therefore restrict our evaluation to the baselines described above while acknowledging prior contributions in this area.

\subsection{Simulation Results}
\begin{table}[t]
\centering
\addtolength{\belowcaptionskip}{-6pt}
\caption{Results on Simulated Data.}
\label{tab:sim_results}
\sisetup{
    reset-text-series = false, 
    text-series-to-math = true, 
    mode=text,
    tight-spacing=true,
    round-mode=places,
    round-precision=2,
    table-format=2.2,
    table-number-alignment=center
}
\begin{tabular}{lccc}
\toprule
      & {PESQ $\uparrow$} & {SI-SDR $\uparrow$} &  {eSTOI $\uparrow$}\\
    \midrule
    No Processing &         1.81     &     2.92     &	    0.59     \\
    \midrule
    SGMSE \cite{richter2023speech}      &  2.63           &      5.37    &	0.70 \\
    MC-SGMSE (ours)      &        2.93   &     6.99     	&   0.73 \\
    AMDM-SE (ours)       &  \textbf{3.11} & \textbf{7.06}  &\textbf{0.75}	 \\
    \bottomrule
\end{tabular}
\end{table}
We first evaluated the performance of the three models trained on the previously described simulated dataset. When using SGMSE \cite{richter2023speech}, we chose channel \#0. 
For both AMDM-SE and MC-SGMSE (multichannel SGMSE, No Attention) we used all four channels.

Table~\ref{tab:sim_results} presents the enhancement performance of all models on the simulated test set, 
These results highlight the effectiveness of the proposed AMDM-SE model. Notably, while both multichannel models significantly outperform the single-channel baseline, AMDM-SE achieves the highest scores across all three metrics. This demonstrates the benefit of incorporating both multichannel input and attention-driven diffusion modeling for speech enhancement in low-reverberation environments. 


\subsection{CHiME-3 Results}

\begin{table}[t]
\centering
\addtolength{\belowcaptionskip}{-4pt}
\addtolength{\abovecaptionskip}{-4pt}
\caption{Experimental results on CHiME3 data.}
\label{tab:CHiME3}
\sisetup{
    reset-text-series = false, 
    text-series-to-math = true, 
    mode=text,
    tight-spacing=true,
    round-mode=places,
    round-precision=3,
    table-format=2.2,
    table-number-alignment=center
}
 \resizebox{\columnwidth}{!}{%
\begin{tabular}{l*{2}{SSS}}
\toprule
    &  {PESQ $\uparrow$} & {eSTOI $\uparrow$} & ovrl$\uparrow$ &
     sig$\uparrow$ & bak$\uparrow$\\
    \midrule
    No Processing &            1.271 &           0.683 &    1.863 &   2.697 &   1.865\\
    \midrule
    CA Dense U-Net \cite{tolooshams2020channel} &           2.436 &     x      & x           &   x & x     \\

    \midrule
    SGMSE \cite{richter2023speech}      &            1.96 &           0.84 &            3.128 &  3.429& \bfseries 4.072 \\
    AMDM-SE (ours) & \bfseries 2.58 & \bfseries 0.91 & \bfseries 3.158 & \bfseries 3.492 & 3.964  \\

    \bottomrule
\end{tabular}
}

\end{table}

The results on the CHiME-3 dataset are presented in Table~\ref{tab:CHiME3}. A substantial improvement is observed when comparing the single-channel baseline to the proposed AMDM-SE model, highlighting the effectiveness of our multichannel, attention-based diffusion approach in more realistic acoustic conditions. SI-SNR results are not reported for the CHiME-3 dataset, as temporal misalignment between the original noisy recordings and the selected target signal (close-talk microphone) renders this metric unreliable in this context. 
%
We also report the PESQ results for the baseline predictive model, CA Dense U-Net \cite{tolooshams2020channel}, using the scores as reported in the original paper.

To facilitate a qualitative comparison, we visualize the speech enhancement outputs in Fig.~\ref{fig:spectrograms}. While both the single-channel SGMSE and the proposed AMDM-SE model effectively suppress most of the background noise, it can be observed that the AMDM-SE exhibits superior reconstruction of speech harmonics. This is particularly evident within the highlighted red frame, where the harmonic structure is preserved with higher fidelity.
\vspace{-0.5em}
\section{Conclusion}
\label{sec:Conclusion}

In this work, we introduced AMDM-SE, an attention-based multichannel diffusion model for speech enhancement. By integrating spatial cues across channels through a unique cross-channel \ac{TF} attention mechanism, our model effectively leverages the rich information available in multichannel recordings. Experimental results on both simulated data and the CHiME-3 dataset demonstrate that AMDM-SE consistently performs well in terms of perceptual quality and intelligibility. These findings underscore the potential of combining generative diffusion frameworks with multichannel spatial modeling for robust speech enhancement in real-world acoustic environments. Incorporating additional multichannel input features, such as \ac{RTF}, offers a promising direction for further improving model performance. 
\vspace{-0.5em}


\label{sec:Acknowledgment}


{
\bibliographystyle{IEEEtran}
\bibliography{refs25}
}







\end{document}